%% file: main.tex
\documentclass[sigconf]{acmart}

\AtBeginDocument{%
  }

\setcopyright{acmlicensed}
\copyrightyear{2026}
\acmYear{2026}
\acmConference[KDD'26 - MLF workshop]{KDD'26 - 9th Workshop on Machine Learning in Finance}{August 9 - 13, 2026}{Jeju, Korea}
\acmISBN{}
\acmDOI{}
\usepackage{balance}
\usepackage{tcolorbox}
\usepackage{listings}
\newtcolorbox{placeholderbox}{
  colback=yellow!15,
  colframe=red!70,
  boxrule=0.5pt,
  arc=2pt,
  left=4pt, right=4pt, top=2pt, bottom=2pt,
  fontupper=\small\itshape,
  before upper={\textbf{TODO:}~}
}

\begin{document}

\title{Detection, Attribution, Narration: An End-to-End Pipeline for Explainable Money Mule Identification}


\author[Y. Zhang]{Yuge Zhang}
\email{yugezhang@ocbc.com}
\affiliation{%
  \institution{OCBC, Singapore}
  \city{}
  \country{}
}

\author[Y. Zhang]{Yuanxing Zhang}
\email{yuanxingzhang@ocbc.com}
\affiliation{%
  \institution{OCBC, Singapore}
  \city{}
  \country{}
}

\author[Y. Jin]{Yichao Jin}
\email{jinyichao@ocbc.com}
\affiliation{%
  \institution{OCBC, Singapore}
  \city{}
  \country{}
}

\author[K. Razis]{Khairul Amsyar Mohd Razis}
\email{khairulmohdrazis@ocbc.com}
\affiliation{%
  \institution{OCBC, Singapore}
  \city{}
  \country{}
}

\author[N. Choo]{Nicholas Qi An Choo}
\email{nicholaschoo3@ocbc.com}
\affiliation{%
  \institution{OCBC, Singapore}
  \city{}
  \country{}
}

\author[K. Wong]{Kai Yin Anders Wong}
\email{anderswong@ocbc.com}
\affiliation{%
  \institution{OCBC, Singapore}
  \city{}
  \country{}
}

\author[X. Tang]{Xinyan Tang}
\email{tangxy@ocbc.com}
\affiliation{%
  \institution{OCBC, Singapore}
  \city{}
  \country{}
}

\author[K. Zhu]{Kenneth Zhu Ke}
\email{kennethzhu@ocbc.com}
\affiliation{%
  \institution{OCBC, Singapore}
  \city{}
  \country{}
}

\author[W. Lee]{Wee Keong Dennis Lee}
\email{dennislee@ocbc.com}
\affiliation{%
  \institution{OCBC, Singapore}
  \city{}
  \country{}
}

\author[J. Zhao]{Jingyuan Zhao}
\email{jingyuanzhao@ocbc.com}
\affiliation{%
  \institution{OCBC, Singapore}
  \city{}
  \country{}
}



\begin{abstract}
Money mule accounts are critical facilitators of financial fraud, yet detecting them at scale remains challenging due to the heterogeneous nature of transactional and behavioural data. We present an end-to-end pipeline for customer-level mule detection comprising three stages: (1)~a LightGBM classifier trained on 280 engineered features spanning transaction patterns, account demographics, network topology, and temporal behaviour; (2)~a TreeSHAP attribution layer that decomposes each prediction into feature contributions; and (3)~a large language model (LLM) module that converts SHAP attributions into analyst-facing natural-language narratives. We evaluate across three open-weight LLM families and assess explanation quality through analyst feedback. In a live production deployment, the system achieves a yield rate of 89\%, up from 61\% under the incumbent rule-based system, with monthly alert volume expanding from 211 to 302, reflecting broader true-positive coverage rather than increased noise. This corresponds to a 60\% incremental adverse detection beyond existing review workflows, substantially outperforming the rule-based approach. Qualitative feedback from analysts indicates that LLM-generated narratives reduce cognitive load during alert triage. We further discuss implications of deploying LLM-augmented explainability in regulated financial environments.
\end{abstract}

\maketitle

\section{Introduction}

Money mules, which are usually individual bank accounts involved in laundering the proceeds of fraud, represent one of the most prevalent drivers in the financial crime ecosystem. Mule accounts receive illicit funds from victims and rapidly disperse them onward, often across jurisdictions, making recovery difficult and enabling downstream criminal activity at scale~\cite{pourhabibi2020fraud}. Despite increasing regulatory scrutiny, detection remains difficult because mule accounts often mimic legitimate customer behaviour during onboarding and early account activity, revealing anomalous patterns only when examined across multiple feature dimensions simultaneously.

A further operational challenge is that detection alone is insufficient. In practice, every flagged customer must be reviewed by a human analyst before action can be taken. Therefore, the usefulness of a detection system depends not only on its precision and recall, but also on how quickly and confidently analysts can triage the alerts it produces. This \emph{explainability} requirement is often underweighted in the academic literature, but is critical in production.

The incumbent approach in most financial institutions relies on deterministic threshold-based rules (e.g., more than $N$ transactions to new payees within $T$ days). While interpretable, these approaches suffer from a well-documented precision-recall trade-off. Tightening rules reduces false positives but allows sophisticated mules to evade detection, while loosening them overwhelms analyst queues with false alerts. The rigidity of static rules also means that they cannot adapt to the evolving mule typologies without costly, repeated manual recalibration.

Critically, these rule-based systems usually operate on individual features in isolation. Money mule behaviour, however, is inherently multi-dimensional, where no single indicator is sufficiently diagnostic. Instead, it relies heavily on the \emph{combination} of subtle anomalies across transaction velocity, network structure, temporal patterns, and account demographics. This distinctive signature cannot be easily captured by deterministic rules.

This paper presents an end-to-end pipeline that addresses both detection and explainability challenges. Our contributions include

\begin{itemize}
\item \textbf{High-dimensional ML classifier:} A LightGBM model \cite{ke2017lightgbm} trained using 280 engineered features for customer-level mule detection, with calibration analysis demonstrating reliable probability estimates across score ranges.

\item \textbf{LLM-augmented explainability:} A pipeline that feeds SHAP \cite{lundberg2017unified} attributions into large language models to generate natural-language narratives. We assess output quality through analyst feedback.

\item \textbf{Live production validation:} Deployment in an AML (Anti-Money Laundering) production environment achieves a yield rate of 89\%, up from 61\% under the incumbent rule-based system. Monthly alert volume expanded from 211 to 302, reflecting broader true-positive coverage rather than increased noise. This corresponds to 60\% incremental adverse detection beyond existing review workflows, with qualitative analyst feedback indicating reduced cognitive load during triage compared to raw SHAP visualizations.

\item \textbf{Operational insights:} Practical lessons on integrating ML and LLM components into analyst workflows within regulated financial environments, including guardrails against LLM hallucination and considerations for model governance.
\end{itemize}



\section{Related Work}
Our work sits at the intersection of three research streams, including machine learning for financial crime detection, post-hoc explainability for high-stakes models, and the emerging use of LLMs to translate quantitative explanations into natural language. We review each in turn, highlighting the gap our pipeline addresses.

\subsection{Machine Learning for AML Detection}

Machine learning has been applied to fraud and AML detection for more than two decades~\cite{pourhabibi2020fraud}, yet work targeting money mule accounts specifically remains limited. Most published approaches frame the problem at the transaction level and rely on graph-based methods. Savage et al.~\cite{savage2017detection} used supervised learning in small networks grouped by transactional interactions. More recently, Jambhrunkar et al.~\cite{jambhrunkar2025muletrack} proposed MuleTrack, a Markov-chain-based framework for detecting mule accounts in India's UPI (Unified Payments Interface) system. Huang et al.~\cite{huang2025enhancing} presented the MuleTrace algorithm to locate mule nodes within transaction graphs by tracking entire laundering chains. 

While graph-based methods are well-suited to network-level detection, they require access to full transaction graphs, which is often infeasible in siloed institutional environments. In contrast, our approach operates on \emph{tabular features engineered at the customer level}, requiring only data available within a single institution. Gradient boosting methods, particularly LightGBM~\cite{ke2017lightgbm}, have demonstrated strong performance on tabular financial data. Taha and
Malebary~\cite{taha2020intelligent} showed that an optimised LightGBM achieves state-of-the-art results on credit card fraud detection. Xu et al.~\cite{xu2023efficient} proposed deep boosting decision trees that embed neural networks into gradient boosting while preserving interpretability. To date, gradient boosting approaches remain the dominant approach in deployed AML systems, yet published work rarely reports production deployment metrics such as analyst yield rate and increased coverage of real-world cases.

\subsection{Explainability in High-Stakes ML}

In regulated financial environments, model transparency is both an operational necessity and a compliance requirement. Regulators, including the United States (US) government \cite{occ2011sr117} and European Union (EU) \cite{euaiact2024}, increasingly mandate financial institutions to demonstrate how model output informs decisions. SHAP (SHapley Additive exPlanations)~\cite{lundberg2017unified} had become the de facto standard for local feature attribution, due to its game-theoretic foundations and the efficiency of the TreeSHAP algorithm~\cite{lundberg2020local} for tree-based models. 

However, a persistent gap exists between what these methods produce, mainly numeric feature-contribution vectors, and what operational analysts need to act on. Waterfall plots and beeswarm diagrams require statistical literacy to interpret and slow down alert triage, particularly when analysts must process hundreds of alerts per day. This \emph{last-mile explainability gap} is the primary motivation for our LLM interpretation module.

\subsection{LLMs as Explanation Narrators}

The use of LLMs to convert quantitative AI model outputs into natural language is a rapidly growing research direction. Martens et al.~\cite{martens2025tell} demonstrated that LLM-generated narratives from SHAP values are found convincing by over 90\% of surveyed non-experts and help users more accurately summarize AI decisions compared to raw SHAP plots alone. Zytek et al.~\cite{zytek2024explingo} developed Explingo, a dual-LLM system comprising a narrator that generates explanation narratives and an automated grader that evaluates
them on accuracy, completeness, and fluency. Bello et al.~\cite{bello2025three} presented a three-level framework for LLM-enhanced explainable AI, including a case study that translates SHAP values from a loan-denial model into conversational narratives. Zeng~\cite{zeng2024enhancing} further explores using LLMs to enhance the interpretability of SHAP values specifically, demonstrating that model-generated narratives can surface feature interactions that raw attribution plots obscure.

These works established the feasibility of LLM-based XAI narration. However, evaluation to date has focused on narrative quality metrics rather than downstream operational impact, and deployment constraints such as on-premise hosting for data residency have not been highlighted. Our work extends this direction with production-validated evidence from a live AML setting using self-hosted open-weight models.

\section{Methodology}

\subsection{System Overview}

\begin{figure*}[t]
  \centering
  \includegraphics[width=\textwidth]{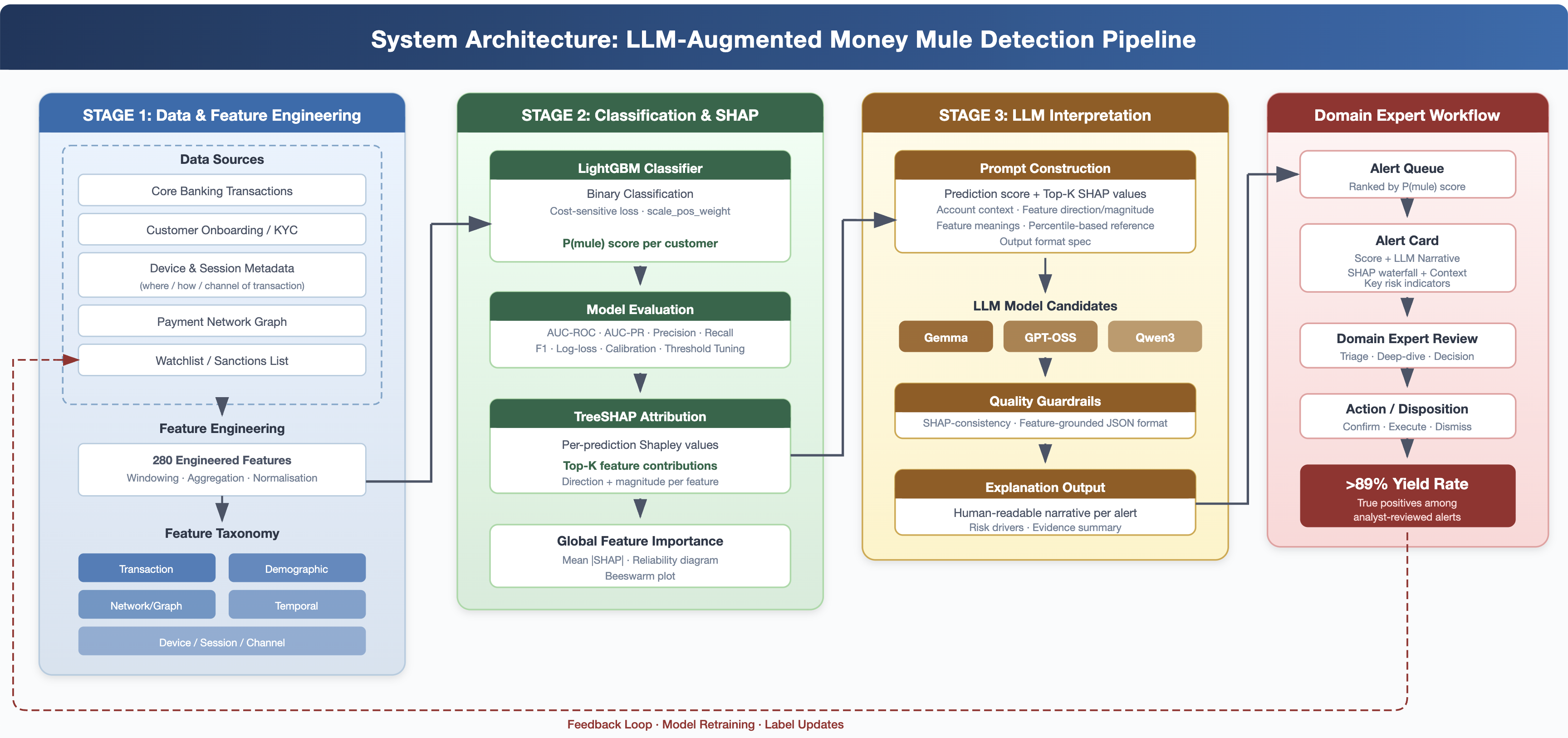}
  \caption{End-to-end system architecture for LLM-augmented money mule detection. The pipeline flows from raw data sources (left) through feature engineering (280 features), LightGBM classification with TreeSHAP attribution, LLM-based narrative generation, and into the domain expert workflow. Dashed line indicates the feedback loop for model retraining.}
  \label{fig:architecture}
\end{figure*}

The pipeline executes three modules sequentially for each customer: (1)~feature engineering and classification, (2)~SHAP attribution, and (3)~LLM interpretation. This is followed by a domain expert workflow for alert triage.  Figure~\ref{fig:architecture} presents the end-to-end architecture. First, raw data from multiple source systems are transformed into a tabular feature matrix at the customer level and scored by a LightGBM classifier. The model outputs are then decomposed via TreeSHAP into per-feature contributions and narrated by self-hosted open-weight LLM models into a domain-expert-facing explanation. Alerts are ranked by predicted probability and surfaced to human analysts together with the generated narrative. A feedback loop from confirmed dispositions back into the watchlist and label pipeline to support periodic model retraining.

\subsection{Data and Feature Engineering}

\textbf{Data sources.} The model ingests data from multiple internal systems, including core banking transactions, customer onboarding records, device and channel metadata, and internal/external watchlist and sanctions feeds.

The training dataset includes 2,176 customer-level records across 2,115 unique customers. 

Labels are derived from two primary sources of ground truth. The first consists of customers subject to Seizure Orders (SOs) filed by police officers under the Criminal Procedure Code (CPC), wherein affected customers are assigned a positive label, reflecting a legal determination that the funds are suspected proceeds of criminal activity. The second source comprises analyst-confirmed money mule cases, where customers have been reviewed and adjudicated by financial crime investigators. Customers confirmed as mules are labeled positive; those cleared are labeled negative. 

In total, 1,680 customers (77.2\%) are labeled as mules (positive class) and 496 (22.8\%) as non-mules (negative class), yielding a positive-to-negative ratio of approximately 3.4:1. This distribution does not reflect the true population prevalence of money mule activity. To better approximate the real-world class distribution, additional negative samples are drawn from the unlabeled population and incorporated into the training set, reducing the positive-to-negative ratio to 1:5.

\textbf{Feature taxonomy.} Over 280 features are engineered at the customer level and organized into five overlapping feature groups, as summarized in Table~\ref{tab:features}.

\begin{table}[t]
\centering
\caption{Summary of Feature Groups}
\label{tab:features}
\begin{tabular}{p{1.5cm}p{0.7cm}p{5.3cm}}
\toprule
\textbf{Feature Group} & \textbf{Count} & \textbf{Representative Examples} \\
\midrule
Transaction patterns & 263 & Transaction count and amount for all (debit and credit), debit, and credit transactions; credit-to-debit ratios; pass-through frequency and amount; equivalent statistics for STR (Suspicious Transaction Report)-linked counterparties and high/medium-risk geographies \\
Account demographics & 15 & Customer tenure, age, country of domicile, number of active accounts, recency of account opening/closing, number of police orders received, and recency of last police order \\
Network / graph & 47 & Number and types of C2C relationships, number of connected customers with STR filings, transaction value linked to network STRs \\
Temporal behaviour & 37 & Short-term to long-term velocity ratios comparing 7-day activity against 180-day baseline for all (debit and credit), debit, credit, and STR-linked transactions, capturing sudden spikes in activity \\
Device / session / channel & 16 & Number of distinct transaction channels used across multiple lookback windows, including channel diversity for high-risk geography transactions and STR-linked entities \\
\bottomrule
\multicolumn{3}{p{0.95\linewidth}}{\footnotesize \textit{Note:} Feature counts sum to more than the total of 280 as certain features span multiple groups.}
\end{tabular}
\end{table}

All features are computed over multiple time windows (e.g., 7, 30, 90, 180~days) to capture both short-term bursts and longer-term behavioural drift.

Features are constructed by aggregating raw transactional and relational data at the 
customer level across multiple lookback windows of 7, 30, 90, and 180 days. For each 
window, aggregation functions including count, sum, min, max, and standard deviation 
are applied to capture transaction frequency, volume, and dispersion. The diversity of counterparties and 
channels is measured using cardinality counts across unique recipients, senders, countries, and transaction channels. Two types of derived ratio features are computed, including credit-to-debit ratios for both transaction count and transaction value within each look-back window, as well as short-to-long-term velocity ratios that contrast 7-day activity with a 180-day baseline to identify sudden increases in transactional behaviour.

Missing values in numeric features are imputed with zero, which is semantically 
appropriate as the absence of a transaction record implies zero activity. Categorical 
features retain missing values and are handled natively by the model. To prevent 
division-by-zero artefacts in ratio features, zero denominators are substituted with a 
missing indicator prior to computation.

No explicit post-engineering feature selection procedure, such as recursive feature 
elimination or correlation-based filtering, was applied. The full set of 280 engineered 
features was presented to the model, with implicit feature selection performed by the 
gradient-boosted tree algorithm through its split-gain optimisation criterion.

\subsection{LightGBM Classifier}

We use LightGBM~\cite{ke2017lightgbm} for binary classification at the customer level, predicting the probability that a customer is operating as a money mule. LightGBM is chosen for its efficiency on high-dimensional tabular data, native categorical feature handling, and strong empirical performance in deployed financial systems.

\textbf{Class imbalance.} Although money mule accounts are rare in the real-world population, we apply negative sampling during training to construct a dataset with a positive-to-negative ratio of 1:5. This introduces an engineered class imbalance in the training set. To address this, we use the \texttt{scale\_pos\_weight} parameter, set to the ratio of negative to positive samples in the training data, which up-weights the loss contribution of positive examples.

\textbf{Hyperparameter tuning.} We perform grid search over the following hyperparameter space, selecting the configuration that maximises AUC-PR on the validation set as shown in table \ref{tab:hyperparameters}.

\textbf{Temporal train/test split.} To prevent data leakage and simulate realistic deployment conditions, we split the data temporally. In particular, the model is trained on customers observed before a cutoff date and evaluated on customers appearing after that date.

\textbf{Calibration.} Since the predicted probability directly determines alert ranking and analyst workload, well-calibrated scores are essential. We assess calibration using reliability diagrams and the Expected Calibration Error (ECE), and apply Platt scaling where necessary. Platt scaling fits a logistic regression on a held-out validation set to map raw LightGBM scores to better-aligned probabilities, ensuring that a predicted score of 0.8 more faithfully reflects an 80\% empirical likelihood of money mule behaviour rather than being an arbitrary model output.

\subsection{SHAP Attribution Layer}

For each customer selected for explanation—either those exceeding the alert threshold or a user-defined top‑N set—we compute exact SHAP values using TreeSHAP~\cite{lundberg2020local}, which exploits the tree structure of LightGBM for polynomial-time computation. TreeSHAP decomposes the predicted log-odds into additive per-feature contributions, satisfying local accuracy, missingness, and consistency axioms~\cite{lundberg2017unified}.

From the full set of 280 SHAP values per prediction, we select the \textbf{top 10 features by absolute SHAP magnitude} for downstream LLM interpretation. This selection balances two concerns, providing sufficient context for a meaningful narrative while keeping the prompt concise enough for reliable LLM generation. For each selected feature, we extract the feature name, its raw value for the account, the SHAP contribution value, and the direction of influence (i.e., positive = increased probability of the mule, negative = decreased probability).

\subsection{LLM Interpretation Module}

The final stage translates the structured SHAP output into a natural-language narrative that analysts can read during alert triage.

\begin{table}[htbp]
\centering
\caption{Selected Hyperparameters for the Model}
\label{tab:hyperparameters}
\begin{tabular}{ll}
\toprule
\textbf{Hyperparameter} & \textbf{Selected Value} \\
\midrule
num\_leaves         & 31   \\
max\_depth          & 8    \\
learning\_rate      & 0.03 \\
min\_child\_samples & 200  \\
subsample           & 0.7  \\
colsample\_bytree   & 0.7  \\
reg\_alpha          & 0.1  \\
reg\_lambda         & 5.0  \\
\bottomrule
\end{tabular}
\end{table}

\textbf{Deployment.} All LLMs are deployed on-premise on a single NVIDIA H100 GPU to ensure that customer data never leaves the institution's infrastructure. We evaluated three models, including Gemma-3-27B \cite{team2024gemma}, GPT-OSS-120B \cite{agarwal2025gpt}, and Qwen3-Next-80B-A3B-Instruct \cite{yang2025qwen3}, with one selected for production deployment based on the evaluation described below. 

\textbf{Prompt design.} The prompt consists of three elements.

\begin{enumerate}
  \item \textbf{Prediction context:} the customer's mule probability score and the score percentile.
  \item \textbf{SHAP evidence:} the top-10 feature attributions, each presented as a (feature name, feature value, SHAP contribution, direction) tuple.
  \item \textbf{Instruction template:} specifying output structure using a fixed JSON schema with red flags, reasons, and evidence references; tone (factual, concise); and constraints (cite only features present in the SHAP evidence; do not speculate beyond the provided data).
\end{enumerate}

\textbf{Narrative constraints and review.} We constrain LLM-generated narratives through the prompt template and periodic human review to reduce the risk of unsupported or misleading explanations.

\begin{itemize}
  \item \textbf{Prompt constraints:} the prompt instructs the LLM to cite only features present in the top-10 SHAP input, avoid speculation beyond the provided case data, and structure the output using a fixed JSON schema. The prompt also requires each red flag to reference the specific feature names supporting the explanation.
  \item \textbf{Human review:} domain experts provide periodic feedback on narrative quality through a lightweight rating mechanism integrated into the alert triage interface, enabling ongoing monitoring of explanation fidelity.
\end{itemize}

\section{Experiments and Results}

\subsection{Experimental Setup}

\textbf{Temporal split.} The dataset is divided by a temporal cutoff. Accounts observed before the cutoff form the training and validation sets, and accounts appearing only after the cutoff date form the held-out test set. 

The dataset is split chronologically into training, validation, and test sets to prevent data leakage. The model is trained on data from January 2025 up to eight weeks before each test period. The eight weeks immediately preceding the test period serve as the validation set to monitor model performance. Testing is conducted over four consecutive weekly periods in December 2025.

\textbf{Evaluation metrics.} We report standard classification metrics where available, including precision, recall, F1-score, and AUC-PR, on the held-out test set. However, the primary metric for evaluating operational impact is
\emph{yield rate}, defined as

\begin{equation}
  \text{Yield Rate} = \frac{\text{Alerts confirmed as true mules}}
    {\text{Total alerts reviewed by analysts}}
\end{equation}

The yield rate is measured on live production alerts rather than the held-out test set, reflecting the end-to-end effectiveness of the system, including the analyst-in-the-loop decision process. This metric directly captures what matters operationally: how often an analyst's time spent reviewing an alert leads to a confirmed case.

\textbf{Baseline.} We compare against the incumbent rule-based detection system that was in production prior to our pipeline. The rule-based system uses deterministic thresholds on individual transaction features (e.g., transaction velocity, counterparty count, in/out ratios) to flag suspicious accounts. This represents the realistic production baseline rather than an academic benchmark.

\subsection{Classification Performance}
Table~\ref{tab:model_comparison} summarises the available classification
metrics for the LightGBM model on the held-out test set.

\begin{table}[h]
\centering
\caption{LightGBM model performance on the held-out test set. Precision, recall, and F1-score are reported at the operational threshold of 0.8.}
\label{tab:model_comparison}
\begin{tabular}{lc}
\toprule
\textbf{Metric} & \textbf{Value} \\
\midrule
Precision   & 0.909 \\
Recall      & 0.816 \\
F1-Score    & 0.860 \\
AUC-ROC     & 0.824 \\
AUC-PR      & 0.943 \\
Log-Loss    & 0.450 \\
\bottomrule
\end{tabular}
\end{table}

\textbf{Threshold selection and model performance.} Each week, the model scores between 20,000 and 30,000 cases, of which analysts can investigate approximately 100. The operating threshold of 0.8 was selected as the F1-optimal cut-off, meaning it is the score above which the balance between precision and recall is maximised. This threshold determines which cases are surfaced as alerts from the full scored population, reducing tens of thousands of cases to a queue that fits within analyst capacity.

\textbf{Precision-recall trade-off.} At threshold 0.8, the model achieves a mean precision of 0.909 and a mean recall of 0.816 across the four December 2025 backtesting weeks. The high precision means that approximately 9 in 10 cases flagged to analysts are genuine money mules, keeping the workload focused and efficient. The recall of 0.816 reflects the inherent cost of setting a high threshold: roughly 1 in 5 true money mule cases score below 0.8 and are not immediately escalated. This trade-off is a deliberate design choice. Setting the threshold lower would recover more true cases but would also push the weekly alert volumes beyond what analysts can handle, diluting the quality of reviews. The F1 score of 0.860 at this threshold represents the best achievable balance between detection coverage and operational precision given the analyst capacity constraint.

It is important to note that the backtesting evaluation was conducted on the reviewed cases available from that a subset period of the full weekly population. This could be interpreted as an estimate of model performance rather than a count of production alerts.

Beyond the backtesting evaluation, the model’s real-world effectiveness is validated through the yield rate. Over the four weeks of December 2025, analysts reviewed 100 model-selected cases per week. Of these, 89\% were confirmed as true money mules, yielding a production precision that directly reflects how useful the model’s prioritisation is in practice.

The yield rate of 89\% means that analysts spent the vast majority of their review capacity on genuine cases, with fewer than 11 in every 100 alerts turning out to be false escalations. This level of precision is operationally meaningful in a context where analyst time is the binding constraint. A high yield rate ensures that the 100-case weekly capacity is not diluted by noise, allowing investigators to focus their effort on the highest-risk cases.

\textbf{Calibration.} The LightGBM model outputs raw probability scores that were subsequently adjusted using Platt scaling, which is a post-hoc calibration technique that fits a logistic regression on a held-out calibration validation set to map raw model scores to better-aligned probabilities. This ensures that a predicted score of 0.8 more faithfully reflects an empirical probability of 80\% of money mule behaviour, rather than being an arbitrary model output.

Calibration quality is assessed using the Expected Calibration Error (ECE), which measures the weighted average gap between predicted scores and observed positive rates across score bins. The combined ECE across all four December 2025 backtesting weeks is 0.079, indicating acceptable calibration overall.

\begin{figure}[t]
    \centering
    \vspace{-0.5em}
    \includegraphics[width=0.65\linewidth]{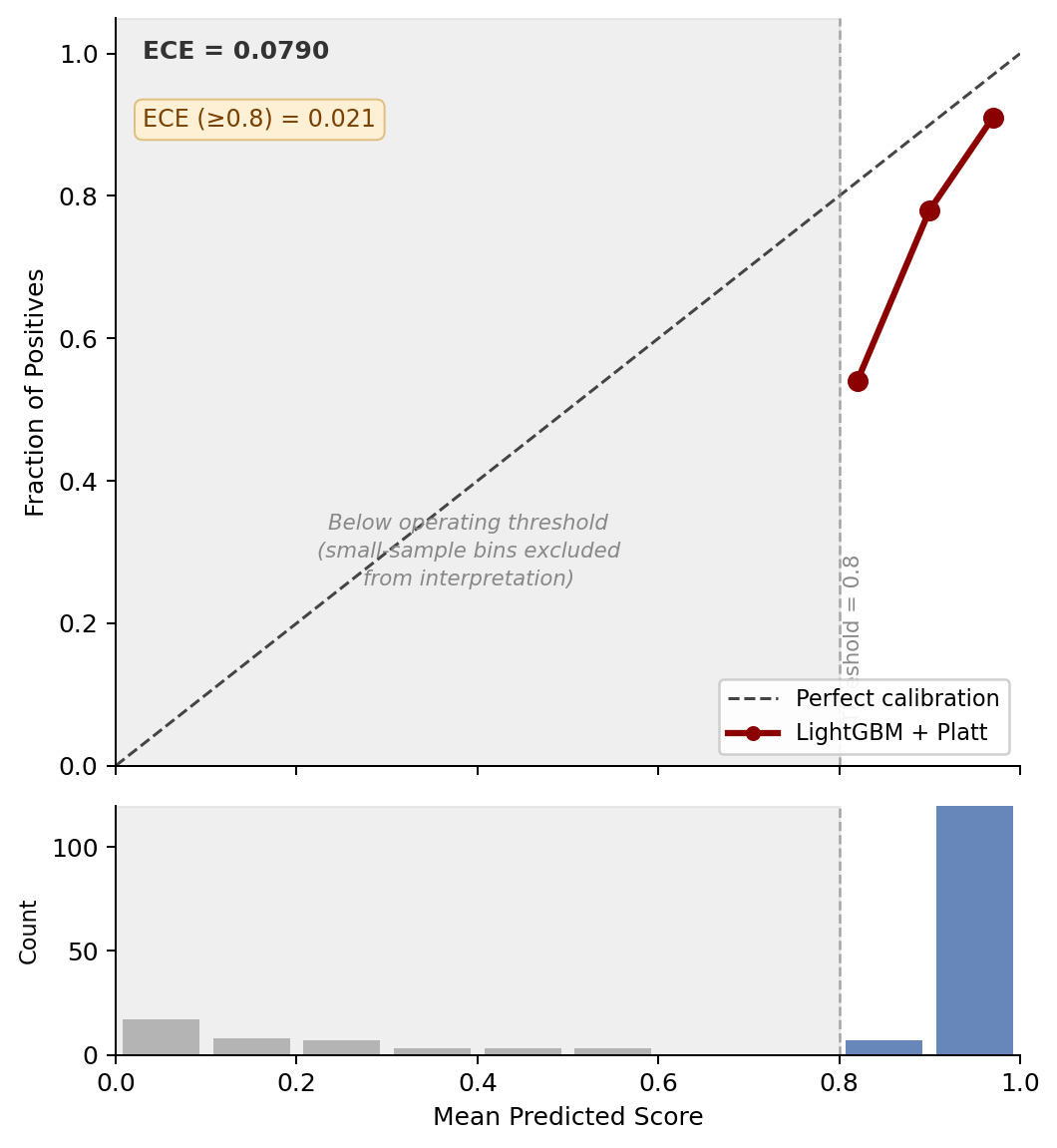}
    \caption{Reliability diagram for the LightGBM + Platt-scaled model across four December 2025 backtesting weeks (combined). The dashed line represents perfect calibration. The shaded region (score $<$ 0.8) contains sparse labelled bins and is excluded from calibration interpretation. The operating threshold is set at 0.8. Sample sizes per bin are annotated for the high-score region (score $\geq$ 0.8). ECE ($\geq$ 0.8) $=$ 0.021; combined ECE (all bins) $=$ 0.079.}
    \label{fig:reliability}
\end{figure}

Figure~\ref{fig:reliability} illustrates the reliability diagram of the model. In the high-score region (score $\geq$ 0.8), where the operating threshold sits at 0.8, the model curve closely tracks the perfect calibration diagonal, with an ECE of 0.021. The zigzag pattern in the lower score range attributes to small bin sample sizes ($n \leq 8$) rather than systematic miscalibration, and falls outside the operationally relevant score range. The $n = 102$ bin at score $\approx$ 0.97, representing the largest and most reliable sample. This aligns closely with the diagonal, further supporting calibration quality at the decision boundary.

\subsection{Feature Importance}
\begin{figure}[t]
    \centering
    \includegraphics[width=0.75\linewidth]{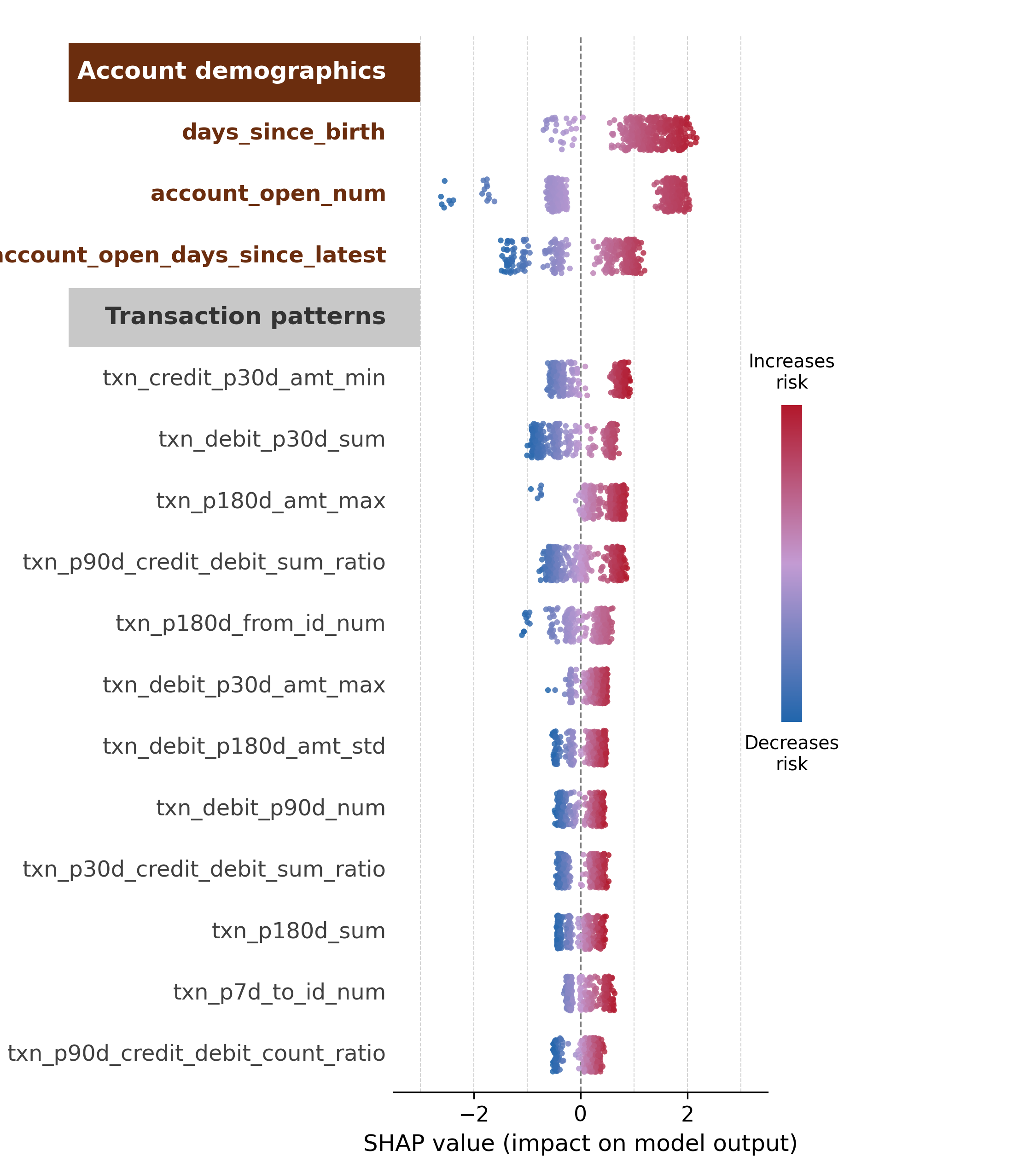}
    \caption{SHAP beeswarm plot of the top 15 features by mean $|\text{SHAP}|$ value. Each point represents one of the 400 scored customers pooled across four December 2025 backtesting weeks (W1-W4). Features are ordered top-to-bottom by decreasing mean absolute SHAP value. Point colour encodes the direction of each feature's contribution. Red indicates the feature value pushes the model prediction toward higher risk, while blue indicates a risk-reducing effect. The $x$-axis represents the SHAP value, with positive values increasing predicted risk and negative values decreasing it. Features are grouped into Account demographics (top 3) and Transaction patterns (ranks 4-15).}
    \label{fig:shap}
\end{figure}

We use mean absolute SHAP values across the test set to assess global feature importance. Figure~\ref{fig:shap} presents a beeswarm plot of the top 15 features ranked by mean $|\text{SHAP}|$, pooled across 400 scored customers from the four December 2025 backtesting weeks, collectively accounting for approximately 46\% of the total mean absolute SHAP mass across all 280 features.

The analysis reveals two dominant feature groups. Account demographic features occupy the top three positions, collectively accounting for 19.5\% of total mean absolute SHAP mass. The top-ranked feature is \mbox{\texttt{days\_since\_birth}} (mean $|\text{SHAP}|$ $=$ 1.31), then \mbox{\texttt{account\_open\_num}} (1.21) and \mbox{\texttt{account\_open\_days\_since\_latest}} (0.81). Directionally, higher values of \texttt{days\_\allowbreak since\_\allowbreak birth} are associated with increased risk, indicating that older customers are disproportionately flagged, potentially reflecting greater vulnerability to recruitment as money mules. \texttt{account\_\allowbreak open\_\allowbreak num} exhibits a bimodal distribution, where both very low and very high values push predictions toward higher risk, suggesting that newly created single-account holders and accounts with an unusually large number of linked accounts are flagged. Transaction pattern features (ranks 4-15) contribute a further 26.5\%, with credit-to-debit ratio features across multiple lookback windows (7, 30, 90, and 180 days) and minimum credit amount features emerging as the most discriminative among them. Notably, lower transaction activity and lower credit amounts consistently shift SHAP values negative, implying that financially inactive accounts are a stronger mule signal than volume-driven behaviour. Although multiple ratio features share similar lookback windows and may exhibit collinearity, SHAP values are computed on model output directly and thus reflect marginal contribution after all feature interactions are accounted for. 

\begin{table*}[!t]
\centering
\caption{LLM model comparison on 20 confirmed money-mule alerts. Rubric dimensions are 1-5 mean scores (higher is better); latency is the mean wall-clock inference time per alert on the same single-H100 on-premise deployment.}
\label{tab:llm_comparison}
\begin{tabular}{lccccc}
\toprule
\textbf{Model} & \textbf{Factual Accuracy} & \textbf{Typology Plausibility} & \textbf{Completeness} & \textbf{Actionability} & \textbf{Latency (s/alert)} \\
\midrule
GPT-OSS-120B       & 4.20 & 3.50 & 2.90 & 3.35 & 5.7 \\
Qwen3-Next-80B-A3B & 3.65 & 3.35 & 3.30 & 3.35 & 20.2 \\
Gemma-3-27B        & 2.55 & 2.75 & 2.80 & 2.60 & 40.4 \\
\bottomrule
\end{tabular}
\end{table*}

\subsection{LLM Explanation Evaluation}
\label{sec:llm_evaluation}

We evaluated three open-weight LLM families, GPT-OSS-120B, Qwen3-Next-80B-A3B, and Gemma-3-27B, on 20 confirmed money-mule alerts under the production prompt template (Appendix~\ref{app:prompts}). Each output was scored on a 1-5 scale (higher is better) across four dimensions by blinded independent evaluators:

\begin{itemize}
  \item \textbf{Factual accuracy:} numeric values, percentile claims, and feature descriptions are faithful to the case input; every cited feature resolves to one present in the top-10 SHAP evidence; no fabricated SHAP values.
  \item \textbf{Typology plausibility:} each flag maps a real data pattern to a recognised mule typology such as pass-through, velocity spikes, credit-to-debit imbalance, high activity in new accounts, counterparty diversity, or dormant reactivation.
  \item \textbf{Completeness:} coverage of the material signals, especially raw feature value exceeds 90th percentile of the training population, that an analyst would expect to see in the narrative.
  \item \textbf{Actionability:} flags are distinct, data-anchored, and suitable for direct use in analyst alert write-up without rewording.
\end{itemize}

Latency was measured separately as the mean wall-clock inference time per alert on the single-H100 on-premise deployment. Table~\ref{tab:llm_comparison} summarises the results.

GPT-OSS-120B was selected for production. It has the highest score for factual accuracy (4.20 vs.\ 3.65 and 2.55) and typology plausibility, an equal score on actionability with Qwen, and is the fastest of the three by 3-7$\times$ despite having the largest parameter count. Two factors explain the speed. First, GPT-OSS and Qwen use a mixture-of-experts architecture that activates only a fraction of their parameters per token, while Gemma uses its full 27B parameters at every step. Second, GPT-OSS produces shorter responses on average, with 2.80 red flags per case versus 3.35 for Qwen and 3.55 for Gemma, so fewer tokens are streamed per alert.

Qwen3-Next-80B-A3B is still a viable fallback. It scores highest on completeness by surfacing more red flags per case, but its factual accuracy is lower. It is more prone to SHAP-value fabrication and percentile-claim errors. Gemma-3-27B was ruled out because its factual accuracy and its $\sim$40~s per-alert latency is uneconomical at operational volumes. Across all three models, the most common failure modes are exaggerated SHAP-value reporting, including occasional 10$\times$ magnitude errors, and misinterpretation of credit-to-debit ratio semantics. These observations motivated stricter prompt constraints and periodic human review of generated narratives.

\subsection{Production Deployment and Yield Rate}

The complete pipeline has been deployed in a production AML environment for 1 month, with alerts surfaced to a team of 3 three full-time analysts. Table~\ref{tab:production} compares the operational performance of the proposed pipeline against the incumbent rule-based system, which consists of 5 reports focused on identifying likely bad-actors based on customer profile, non-financial and financial transaction activity, and financial transaction network behaviour. The rule-based metrics are blended across all 5 reports.

\begin{table}[h]
\caption{Operational comparison: proposed pipeline vs.\ incumbent rule-based system (metrics blended across 5 reports) on live production alerts.}
\label{tab:production}
{\setlength{\tabcolsep}{1pt}
\begin{tabular}{lcc}
\toprule
\textbf{Metric} &
\begin{tabular}[t]{@{}c@{}}\textbf{Rule-Based}\\\textbf{(blended across 5 reports)}\end{tabular} &
\textbf{Proposed} \\
\midrule
Yield Rate           & 61\% & 89\% \\
Monthly Alert Volume & 211  & 302  \\
\bottomrule
\end{tabular}
}
\end{table}

The yield rate improvement from 61\% to over 89\% represents a substantial reduction in false positives surfaced to analysts, allowing the team to focus investigative effort on genuinely suspicious accounts. The concurrent rise in monthly alert volume from 211 to 302 reflects broader coverage rather than noise. This indicates that the model surfaces more true positives that the rule-based system had previously missed, without proportionally increasing analyst burden given the higher yield.

\textbf{Analyst feedback on LLM narratives.} Appendix~\ref{app:llm_output} shows two representative LLM outputs alongside the analyst verdicts they received. Through structured feedback sessions, analysts reported that LLM-generated narratives meaningfully reduced the cognitive
  effort required during alert triage. Specifically, analysts noted that narratives allowed them to quickly assess whether a prediction was plausible before committing to a full investigation, and that the structured format (red flags, supporting reasons, and evidence references) aligned well with their existing workflow. Representative analyst feedback highlighted two particularly valued capabilities. First, analysts noted that the narratives \textit{``provide clear guidance into why an alert is classified as higher risk rather than typical customer behaviour and draw attention to specific non-financial red flags that we may not typically monitor.''} Second, analysts identified a proactive monitoring use case: the LLM \textit{``has the capability to help filter higher-risk alerts with low transaction values''}, cases where suspicious behavioural signals are present but adverse transactions have not yet materialised. Rather than dismissing these as low-priority, analysts noted that such alerts can be placed under short-term internal monitoring (typically a few weeks), as flagged accounts frequently exhibit adverse activity within that window. This suggests the system supports not only point-in-time triage but also early-warning identification of emerging money mule activity. 

\section{Discussion}

\subsection{Why ML Outperforms Rules}

The improvement in yield rate from 61\% to over 89\% reflects a fundamental difference in how the two approaches model the behavior of the mule. Rule-based systems operate on individual features in isolation: a single rule might flag accounts exceeding a velocity threshold, while another targets unusual counterparty counts. Each rule captures one behavioural dimension, and combining rules multiplicatively creates rigid decision boundaries that are brittle to slight variations in mule tactics.

The LightGBM model, by contrast, learns non-linear interactions across hundreds of features simultaneously. Money mule behaviour is inherently multi-dimensional, where no single feature is diagnostic, but the combination of moderate anomalies across transaction velocity, temporal patterns, and counterparty diversity creates a distinctive signature. Gradient boosting captures these joint effects naturally through its sequential tree-building process, without requiring analysts to manually specify interaction rules.

A practical consequence is that the ML pipeline also produces higher-quality alerts at a greater volume, giving analysts more cases to review without sacrificing precision. Critically, when analyst capacity increases, the alert volume can be scaled up simply with only marginal loss in yield rate. Rule-based systems lack this flexibility, where tightening one rule often requires loosening another to maintain coverage. Consequently, it is difficult to scale detection without disproportionately inflating false positives.

\subsection{The Value of LLM-Generated Narratives}

A key observation from deployment is that classification performance alone was insufficient to drive analyst adoption. Early iterations of the pipeline surfaced SHAP waterfall plots alongside the mule probability score, but analyst feedback indicated that interpreting raw feature attributions was time-consuming and required familiarity with the model's feature definitions.

The addition of LLM-generated narratives changed the interaction dynamic. Rather than synthesising meaning from a bar chart of 10 numeric values, analysts could read a concise summary that contextualised the risk factors in domain-relevant language (e.g., ``a sudden spike in inflows relative to the account's longer-term baseline'' rather than ``\texttt{txn\_credit\_p7d\_p180d\_ratio\_sum}: SHAP = +0.34''). Analysts reported that this reduced the time to form an initial hypothesis about if an alert warranted deeper investigation.

This finding aligns with recent work on XAI narratives. Cedro and Martens~\cite{cedro2025graphxain} found that over 90\% of non-experts rated SHAP-based narratives as convincing, and Zytek et al.~\cite{zytek2024explingo} showed user preference for LLM narratives over SHAP plots. Our contribution extends these findings to a production AML environment, where the stakes and the scrutiny are considerably higher.

\subsection{Limitations}
We identify a few limitations of the current work as follows.
\begin{itemize}
  \item \textbf{Single-institution evaluation.} All results are from one financial institution. Mule typologies, transaction patterns, and label distributions vary across geographies and banking products, and the pipeline's performance may not transfer directly without retraining and feature adaptation.
  \item \textbf{Label quality.} Ground truth is constructed from police force filings and human analyst confirmations. Both carry inherent limitations. Police filings may be filed with a delay, and analysts may disagree on borderline cases. Furthermore, undetected mules represent survivorship bias in the negative class. The yield rate metric partially mitigates this by measuring confirmed outcomes, but underlying label noise remains.
  \item \textbf{LLM hallucination residual risk.} Despite prompt constraints and periodic human review, LLM-generated narratives can occasionally produce plausible-sounding statements that subtly misrepresent the underlying SHAP evidence. In a compliance context, even infrequent hallucinations carry reputational and regulatory risk. Ongoing monitoring and periodic manual audits are essential.
\end{itemize}

\subsection{Ethical Considerations}
Deploying ML for financial crime detection carries fairness and accountability risks. The model must not disproportionately flag customers based on protected or sensitive characteristics. The feature set excludes gender and ethnicity, but it does include age-related and geography-related variables, such as customer age, country of domicile, and transaction geographies. These variables may be relevant for financial crime risk assessment, but they can also introduce proxy-discrimination risks. We therefore treat demographic and geography-linked features as requiring explicit governance, monitoring, and periodic review rather than assuming that their inclusion is automatically fair or risk-free.

The LLM interpretation layer introduces an additional fairness and conduct risk. Generated narratives may influence analyst judgment even when they are intended only as explanatory aids. To mitigate this risk, the prompt template constrains the LLM to cite only SHAP-supported evidence from the case input and to avoid unsupported speculation. In addition, generated narratives are subject to periodic human review for factual consistency, discriminatory framing, and inappropriate escalation language. The LLM output is not treated as an independent basis for adverse action; final decisions remain with trained financial-crime analysts following the institution's standard review process.

\section{Conclusions}

This paper presents an end-to-end pipeline for money mule detection that combines account-level classification, feature attribution, and LLM-generated natural-language explanations. A LightGBM model trained on 280 engineered features produces calibrated mule probability scores. TreeSHAP decomposes each prediction into per-feature contributions. A self-hosted LLM translates the top-10 attributions into analyst-facing narratives under structured prompt constraints. In live production deployment, the pipeline achieves a yield rate of 89\%, up from 61\% under the incumbent rule-based system. Monthly alert volume expanded from 211 to 302, reflecting broader true-positive coverage. This corresponds to 60\% incremental adverse detection beyond existing review workflows, substantially outperforming the incumbent rule-based approaches. Analyst feedback confirms that LLM-generated narratives reduce cognitive load during triage compared to raw SHAP visualisations.

Three findings may be of broader interest to practitioners deploying ML in regulated financial environments. First, the operational bottleneck in AML detection is often not model accuracy but \emph{explanation usability}. Thus, the LLM interpretation layer was critical to analyst adoption. Second, open-weight models deployed on-premise can satisfy data residency requirements without sacrificing explanation quality, making this approach viable for institutions that cannot use external API-based services. Third, structured prompt templates and periodic human review offer a practical path to reducing hallucination risk in high-stakes settings, although residual risk remains and requires ongoing monitoring.

\balance





\bibliographystyle{ACM-Reference-Format}
\bibliography{reference}

\appendix
\input{appendix_prompts}
\input{appendix_llm_output}




\end{document}

%% file: appendix_prompts.tex
\appendix

\section{Prompt Template}
\label{app:prompts}

This appendix reproduces the prompt used for production deployment and for the LLM evaluation reported in Section~\ref{sec:llm_evaluation}. The prompt consists of a system message (Listing~\ref{lst:system_prompt}) that defines the analyst role, input fields, and required JSON schema, followed by a user message (Listing~\ref{lst:zero_shot}) that injects the per-case SHAP attribution payload at the position marked \texttt{<CASE\_JSON>}.

\lstset{
  basicstyle=\ttfamily\footnotesize,
  breaklines=true,
  breakatwhitespace=true,
  frame=single,
  framesep=4pt,
  captionpos=b,
  numbers=none,
  columns=fullflexible,
  xleftmargin=0pt,
  xrightmargin=0pt,
  showstringspaces=false,
  keepspaces=true,
  aboveskip=6pt,
  belowskip=6pt,
}

\begin{lstlisting}[caption={System prompt.}, label={lst:system_prompt}]
You are a financial crime analyst specializing in detecting Money Mule activities. Your task is to analyze individual cases by examining:
- Feature values (value_raw): Metrics describing transaction behavior or customer profiles, along with their corresponding 90th percentile (ref_p90, above_p90) and 50th percentile (ref_p50) benchmarks across all cases.
- Feature meanings (feature_meaning): Clear explanations of what each feature name (name) represents.
- SHAP values (shap) and absolute SHAP values (abs_shap): Indicators of each feature's impact on the model's prediction for the case.
- Overall risk score and percentile (model_score and score_percentile): The case's risk assessment and its relative standing among all evaluated cases.

Based on this information, identify potential suspicious patterns or red flags that explain why the case has its current risk score. Please explain your findings in plain language, avoiding technical jargon, so that analysts without specialized technical knowledge can easily understand. Please also keep it simple.

Please respond strictly in **valid JSON** following exactly this schema:

{
  "case_id": "<string>",
  "red_flags": [
    {
      "potential_pattern": "<short description of suspicious pattern>",
      "reason": "<detailed reason linking features and SHAP values to the issue>",
      "evidence_references": ["<feature_name>", "..."]
    }
  ]
}
\end{lstlisting}

\begin{lstlisting}[caption={User message.}, label={lst:zero_shot}]
Analyze the following case:
Input data for a case:
<CASE_JSON>

Output (JSON format only):
\end{lstlisting}

%% file: appendix_llm_output.tex
\section{Example LLM Outputs and Analyst Reviews}
\label{app:llm_output}

This appendix presents two representative LLM-generated red flag narratives produced by the production pipeline (Section~\ref{sec:llm_evaluation}), together with the corresponding analyst verdicts recorded after manual review. For readability, we present the narrative content extracted from the JSON outputs rather than the full raw JSON objects. In both cases the analyst confirmed the alert as adverse and classified it under the money-mule typology (MM/UML).

\begin{lstlisting}[caption={Example 1 --- LLM-generated red flag narrative.}, label={lst:llm_output_1}]
The ratio of transactions in the past 7 days compared to the past 180 days is significantly higher than typical, and the total transaction value in the past 7 days is also elevated. This suggests a recent increase in activity that could be suspicious. The customer holds a relatively high number of open accounts (5), and the entity has been around for a considerable time (over 32 years). This, combined with the recent activity surge, could indicate an attempt to use established accounts for illicit purposes. The days since the last account opening is relatively short, which could indicate a sudden increase in account activity. This, combined with the recent surge in transaction activity, could be a red flag.
\end{lstlisting}


\begin{lstlisting}[caption={Example 2 --- LLM-generated red flag narrative.}, label={lst:llm_output_2}]
The model flags this case as very high risk (99.44 percentile). This is driven by a combination of factors. The customer receives money from a very high number of different people (97 counterparties in the last 180 days) and sends money to a larger than typical number of countries (9 countries in the last 180 days). The number of open accounts is also a contributing factor. This, combined with a relatively low total transaction value and the age of the account, suggests potential money mule activity.
\end{lstlisting}


%% file: reference.bib
@String{Springer = "Springer-Verlag" }

@article{pourhabibi2020fraud,
  title={Fraud detection: A systematic literature review of graph-based anomaly detection approaches},
  author={Pourhabibi, Tahereh and Ong, Kok-Leong and Kam, Booi H and Boo, Yee Ling},
  journal={Decision Support Systems},
  volume={133},
  pages={113303},
  year={2020},
  publisher={Elsevier}
}

@article{ke2017lightgbm,
  title={Lightgbm: A highly efficient gradient boosting decision tree},
  author={Ke, Guolin and Meng, Qi and Finley, Thomas and Wang, Taifeng and Chen, Wei and Ma, Weidong and Ye, Qiwei and Liu, Tie-Yan},
  journal={Advances in neural information processing systems},
  volume={30},
  year={2017}
}

@article{lundberg2017unified,
  title={A unified approach to interpreting model predictions},
  author={Lundberg, Scott M and Lee, Su-In},
  journal={Advances in neural information processing systems},
  volume={30},
  year={2017}
}

@inproceedings{savage2017detection,
  title={Detection of Money Laundering Groups: Supervised Learning on Small Networks.},
  author={Savage, David and Wang, Qingmai and Zhang, Xiuzhen and Chou, Pauline and Yu, Xinghuo},
  booktitle={AAAI Workshops},
  pages={5--11},
  year={2017}
}

@inproceedings{jambhrunkar2025muletrack,
  title={MuleTrack: A Lightweight Temporal Learning Framework for Money Mule Detection in Digital Payments},
  author={Jambhrunkar, Ganesh and Sharma, Harsh and Singla, Saurav and Kailasam, Thirumalai},
  booktitle={International Work-Conference on Artificial Neural Networks},
  pages={384--397},
  year={2025},
  organization={Springer}
}

@inproceedings{huang2025enhancing,
  title={Enhancing Anti-Money Laundering by Money Mules Detection on Transaction Graphs},
  author={Huang, Zhenfeng},
  booktitle={Proceedings of the 2025 International Conference on Generative Artificial Intelligence for Business},
  pages={83--88},
  year={2025}
}

@article{lundberg2020local,
  title={From local explanations to global understanding with explainable AI for trees},
  author={Lundberg, Scott M and Erion, Gabriel and Chen, Hugh and DeGrave, Alex and Prutkin, Jordan M and Nair, Bala and Katz, Ronit and Himmelfarb, Jonathan and Bansal, Nisha and Lee, Su-In},
  journal={Nature machine intelligence},
  volume={2},
  number={1},
  pages={56--67},
  year={2020},
  publisher={Nature Publishing Group UK London}
}

@article{martens2025tell,
  title={Tell me a story! Narrative-driven XAI with Large Language Models},
  author={Martens, David and Hinns, James and Dams, Camille and Vergouwen, Mark and Evgeniou, Theodoros},
  journal={Decision Support Systems},
  volume={191},
  pages={114402},
  year={2025},
  publisher={Elsevier}
}

@article{zeng2024enhancing,
  title={Enhancing the interpretability of SHAP values using large language models},
  author={Zeng, Xianlong},
  journal={arXiv preprint arXiv:2409.00079},
  year={2024}
}

@inproceedings{zytek2024explingo,
  title={Explingo: Explaining ai predictions using large language models},
  author={Zytek, Alexandra and Pido, Sara and Alnegheimish, Sarah and Berti-Equille, Laure and Veeramachaneni, Kalyan},
  booktitle={2024 IEEE International Conference on Big Data (BigData)},
  pages={1197--1208},
  year={2024},
  organization={IEEE}
}

@article{bello2025three,
  title={A three-Level Framework for LLM-Enhanced eXplainable AI: From technical explanations to natural language},
  author={Bello, Marilyn and Bello, Rafael and Garc{\'\i}a, Mar{\'\i}a-Matilde and Now{\'e}, Ann and Sevillano-Garc{\'\i}a, Iv{\'a}n and Herrera, Francisco},
  journal={Information Systems Frontiers},
  pages={1--22},
  year={2025},
  publisher={Springer}
}

@article{taha2020intelligent,
  title={An intelligent approach to credit card fraud detection using an optimized light gradient boosting machine},
  author={Taha, Altyeb Altaher and Malebary, Sharaf Jameel},
  journal={IEEE access},
  volume={8},
  pages={25579--25587},
  year={2020},
  publisher={IEEE}
}

@article{xu2023efficient,
  title={Efficient fraud detection using deep boosting decision trees},
  author={Xu, Biao and Wang, Yao and Liao, Xiuwu and Wang, Kaidong},
  journal={Decision Support Systems},
  volume={175},
  pages={114037},
  year={2023},
  publisher={Elsevier}
}

@misc{occ2011sr117,
  title     = {Supervisory Guidance on Model Risk Management},
  author    = {{Office of the Comptroller of the Currency}},
  year      = {2011},
  url       = {https://www.occ.gov/news-issuances/bulletins/2011/bulletin-2011-12a.pdf},
  note      = {Board of Governors of the Federal Reserve System}
}

@misc{euaiact2024,
  title     = {Regulation ({EU}) 2024/1689},
  author    = {{European Parliament and Council of the European Union}},
  year      = {2024},
  url       = {https://eur-lex.europa.eu/eli/reg/2024/1689/oj},
  note      = {Official Journal of the European Union}
}

@article{team2024gemma,
  title={Gemma: Open models based on gemini research and technology},
  author={Team, Gemma and Mesnard, Thomas and Hardin, Cassidy and Dadashi, Robert and Bhupatiraju, Surya and Pathak, Shreya and Sifre, Laurent and Rivi{\`e}re, Morgane and Kale, Mihir Sanjay and Love, Juliette and others},
  journal={arXiv preprint arXiv:2403.08295},
  year={2024}
}

@article{yang2025qwen3,
  title={Qwen3 technical report},
  author={Yang, An and Li, Anfeng and Yang, Baosong and Zhang, Beichen and Hui, Binyuan and Zheng, Bo and Yu, Bowen and Gao, Chang and Huang, Chengen and Lv, Chenxu and others},
  journal={arXiv preprint arXiv:2505.09388},
  year={2025}
}

@article{agarwal2025gpt,
  title={gpt-oss-120b \& gpt-oss-20b model card},
  author={Agarwal, Sandhini and Ahmad, Lama and Ai, Jason and Altman, Sam and Applebaum, Andy and Arbus, Edwin and Arora, Rahul K and Bai, Yu and Baker, Bowen and Bao, Haiming and others},
  journal={arXiv preprint arXiv:2508.10925},
  year={2025}
}

@inproceedings{cedro2025graphxain,
  title={GraphXAIN: narratives to explain graph neural networks},
  author={Cedro, Mateusz and Martens, David},
  booktitle={World Conference on Explainable Artificial Intelligence},
  pages={91--114},
  year={2025},
  organization={Springer}
}
